\documentclass{jaa}
\usepackage{natbib,hyperref}
\usepackage{graphicx}

\begin{document}\sloppy

%%paper title
%%For line breaks \\ can be used within title
\title{Exploring the collinear Lagrangian points of exoplanet systems \\  with P-R drag and oblateness.}

%%author names are separated by comma (,)
%%use \and before the last author name
%%use a * along with the number separated by comma
%for the  author for correspondence
%%\textsuperscript{number} is used for affiliation
%%\affilOne, \affilTwo etc., upto \affilTwentyfive is possible
%%Please note the first letter after \affil is capitalised in the command
%%

\author{Ibtisam Shaikh\textsuperscript{1,*}, Priya Hasan\textsuperscript{2} and S.N.Hasan\textsuperscript{1}}
\affilOne{\textsuperscript{1}Department of Mathematics, Maulana Azad National Urdu University, Hyderabad, 500032, India\\}
\affilTwo{\textsuperscript{2}Department of Physics, Maulana Azad National Urdu University, Hyderabad, 500032, India}

%%escape two column mode for title, affiliation and abstract
%%by giving \twocolumn command as shown

\onecolumn{

\maketitle

%%include \corres to print the corresponding author Email id
\corres{ibzyshaikh@gmail.com}

%%include \msinfo for
%%manuscript information such as
%%received, revised and accepted dates
%%
\msinfo{1 January 2015}{1 January 2015}

%%abstract
\begin{abstract}
In this paper we study the location and stability of the collinear Lagrangian points for the RTBP for the case in which one of the primary bodies is radiating and the other is oblate. We consider the effect of Poynting-Roberson drag and investigate how the location and stability of the Lagrangian points change with changes in the radiation parameter $\beta$ and oblateness $a$.  We apply our results to ten exoplanet systems --  CoRoT-2~b, TOI-1278~b,  HAT-P-20~b, Kepler-75~b,  WASP- 89~b, TIC 172900988~b, 
NGTS 9~b,  LP 714 - 47~b,  WASP- 162~b and  XO- 3~b, data of which has been taken from NASA exoplanet archives, with the aim of finding locations in these exoplanet systems where one can detect asteroids, primodial material or seeds where planet formation can take place. We find that the location of the collinear Lagrangian points changes with variation in radiation pressure and oblateness. Further, for all the ten planetary systems, studied in this paper, the Lagrangian points are unstable and can be locations where we expect to find minor planets, asteroids or debris. The unstability of the the Lagrangian points can be a possible cause of relocation and migration of planetesimals.
\end{abstract}

%%insert keywords separated by 3 hyphens using \keywords{words}

%%close the twocolumn escape here

%%include \doinum{number}for the DOI number in the header
%%include \volnum{number} for the volume number in the header
%%include \year{yyyy} for  year of publication in the header
%%include \pgrange{num--num} page range of article in the header
%%include \artcitid{num} for the article citation id
%%include \lp to print last page of the article
%%include \setcounter{page}{pagenum} for the exact starting page of the article

\doinum{12.3456/s78910-011-012-3}
\artcitid{\#\#\#\#}
\volnum{000}
\year{0000}
\pgrange{1--}
\setcounter{page}{1}
\lp{1}

\section{Introduction}

The three body problem has been an area of active research for over four centuries.
 The restricted three-body problem (RTBP) is a special case  wherein one of the bodies has an infinitesimal mass. 
Further, if one of the primaries is radiating then the motion of the infinitesimal mass in the Restricted Three Body system is affected by perturbations such as radiation pressure, Poynting-Robertson drag (P-R drag), and solar wind drag. 

\cite{1996KosIs..34..159K} studied the photogravitational restricted elliptic three body problem and obtained the locations of the collinear points and studied their stability to the first approximation and derived the condition of stability of the Collinear Lagrange Equilibrium Points (CLEPs).  \cite{1996CosRe..34..411K}  investigated the Photograviatational Hill's problem and found a new class of periodic solution. \cite{subbarao1975note} studied the problem and found that the CLEPs are unstable and Triangular Lagrange equilibrium points (TLEPs) are conditionally stable.  \cite{1980ApJ...238..337S} found that the $L_4$ and $L_5$ points are unstable on a time scale and also discussed the implications of space colonization and a mechanism for producing azimuthal asymmetries in the interplanetary dust complex. In addition, if one of the primaries is oblate, then its oblateness also affects the motion of the other two bodies. These perturbations influence the stability of the Lagrange equilibrium points.   \cite{AbdulRaheem_2006} considered the Coriolis and the Centrifugal forces, together with the effects of oblateness and radiation pressures of the primaries and found that the Coriolis force has a stabilizing tendency and radiation pressure force and oblateness have a destabilizing tendency. 
The stability of the  Lagrange equilibrium points has been studied by \cite{arantza2020stationary} they considered the oblate primary and studied the effects of its oblateness on the location and stability of the equilibrium points. They found the orbits around the equilibrium points and found that the eccentricity of orbits around $L_1$ and $L_3$ increases but around  $L_2$ it decreases with the addition of the oblateness, \cite{ MURRAY1994465} studied the effects of drag in the circular RTBP along with the location and stability of Lagrange equilibrium points.
\cite{article} also found the stationary solutions for the RTBP in case of more massive oblate spheroid. 

In this paper, we obtain the equations of motion for a special case of the RTBP, where the more massive primary is radiating, we take into account the radiation pressure and the  P-R drag and the less massive primary is oblate. It is a non-conservative system due to the Poynting-Robertson drag and solar wind drag.
 There is a constant interplay between the three forces: gravity, P–R drag and radiation pressure,  P–R drag causes dust orbiting a star to lose angular momentum, and spiral slowly into the star, while radiation pressure is an outward force and gravity is attractive. The oblateness also affects the motion of the infinitesimal mass. We obtain the Jacobian, plot the zero velocity curves for different values of the parameters {\it{viz.}}  $\mu$ -- the mass of the primary, $\beta$ -- the radiation parameter,  and $a$ -- the oblateness parameter and find the locations and stability of the CLEPs. Further, we apply our results to ten exoplanet systems namely  CoRoT-2b, TOI-1278 b,  HAT-P-20 b, Kepler-75 b,  WASP- 89 b, TIC 172900988 b, 
 NGTS 9b ,  LP 714 - 47 b,  WASP- 162 b and  XO- 3 b data of which has been taken from  the NASA exoplanet archive\footnote{\url{https://exoplanetarchive.ipac.caltech.edu/}}

%The ratio of the force of radiation to the force of gravity is given by:

%  $$\beta = \frac{\rm{radiation~force}}{ \rm{gravitational~ force}} = \frac{SAQ_{PR} r_1^2}{cGmm_1}$$ 

This case is relevant, particularly for planetary systems in formation where these conditions apply. In this case, the radiation from the protostar varies and the planet in  formation is oblate. The infinitesimal mass is used to describe the accreting dust particles which are the seeds for planet formation. 

%The oblateness parameter is given by:
%$$a = \frac{polar~radius}{equatorial~radius}$$

\section{Formulation of the problem}

The equation of motion of a particle with mass $m$, cross-sectional area $A$, moving under the gravitational influence of two primary bodies, one of which is radiating and the other is an oblate spheroid coupled with the radiation and P-R drag forces is given by  \cite{1979Icar...40....1B} as follows:

\begin{equation}
    m\Ddot{\Vec{r}} = -\frac{Gmm_1}{r_1^3} \Vec{r_1} -\frac{aGmm_2}{r_2 ^3}\Vec{r_2} +\bigg(\frac{SA}{c}\bigg) Q_{PR}\Bigg[\bigg(1-(1+sw)\frac{\dot{\Vec{r}}.\vec{r_1}}{cr_1}\bigg)\hat{r_1}- (1+sw)\frac{\dot{\Vec{r}}}{c}\Bigg]
    \label{1}
\end{equation}
 where $m_1$ and $m_2$ are the masses of the primary bodies and $r_1 $ and $r_2$ are the distances between the particle and $m_1$ and $m_2$ respectively,  $\hat{r_1}$  is unit position vector of particle with respect to $m_1$. $S$, $c$, $Q_{PR}$ are solar energy flux density, speed of light and the radiation pressure force coefficient respectively and $sw$  represents ratio of solar wind drag to  PR  to drag.  We assume $sw$ to be $0.35$\\ 
  Now we introduce a dimensionless quantity $ \beta $ as described in \cite{1995Icar..116..186L}:
  \begin{equation}
   \beta = \frac{\rm{radiation~pressure~force}}{\rm{ gravitational~force}}
         = \frac{SA\ Q_{PR}\ r_1^2}{c\ Gmm_1}
         \label{2}
  \end{equation}
  and $a$ as the oblateness coefficient\\
\begin{equation}
   a = \frac{\rm polar~radius}{\rm equatorial~radius}~{\rm where}~ 
   (a \leq 1)
\end{equation}
We choose dimensions such that $m_1 + m_2 =1$ and $G= 1$. We assume $m_2= \mu$, and $m_1= 1-\mu$, (where $\mu \leq 1/2$) and denote the position of the infinitesimal mass by
  $r= (\xi,\eta,\zeta)$.
 
\begin{figure}[htb]
\scalebox{0.5}[0.5]{\includegraphics{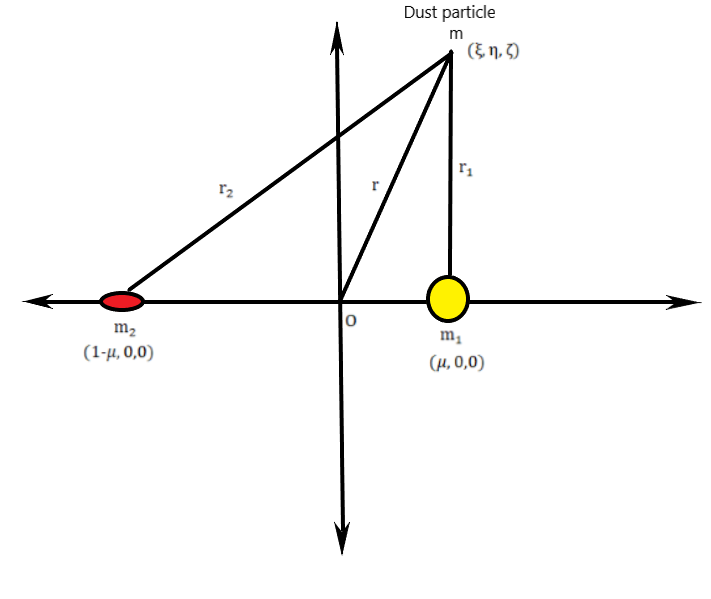}}\centering
    \caption{The origin is denoted by O, the centre of mass of  $m_1$ and $m_2$, the coordinates of $m_1=(\mu,0,0)$ and $m_2=(1-\mu,0,0)$}
    \label{F1}
\end{figure}
Substituting eq.~(\ref{2}) in eq.~(\ref{1}) and resolving  component-wise we get equations of motion of the infinitesimal mass in the inertial frame as follows:\\
\begin{equation}
\begin{split}
    \Ddot{\xi} = -\frac{(1-\beta)(1-\mu)(\xi - \xi_1)}{r_1^3} - \frac{a\mu(\xi-\xi_2)}{r_2^3} +(1+sw) F_{PR,\xi}\\
    \\
    \Ddot{\eta} = -\frac{(1-\beta)(1-\mu)(\eta - \eta_1)}{r_1^3} - \frac{a\mu(\eta-\eta_2)}{r_2^3} +(1+sw) F_{PR,\eta}\\
    \\
    \Ddot{\zeta} = -\frac{(1-\beta)(1-\mu)(\zeta - \zeta_1)}{r_1^3} - \frac{a\mu(\zeta-\zeta_2)}{r_2^3} +(1+sw) F_{PR,\zeta}\\
    \end{split}
    \label{3}
\end{equation}
where
\begin{equation}
    \begin{split}
        F_{PR,\xi} = -\frac{\beta (1-\mu)}{cr_1^2} \Bigg[\frac{\big[(\xi-\xi_1)\dot{\xi}+(\eta-\eta_1)\dot{\eta}+(\zeta-\zeta_1)\dot{\zeta}\big](\xi-\xi_1)}{r_1^2} + \dot{\xi}\Bigg]\\
        \\
        F_{PR,\eta} = -\frac{\beta (1-\mu)}{cr_1^2} \Bigg[\frac{\big[(\xi-\xi_1)\dot{\xi}+(\eta-\eta_1)\dot{\eta}+(\zeta-\zeta_1)\dot{\zeta}\big](\eta-\eta_1)}{r_1^2}+ \dot{\eta}\Bigg]\\
        \\
        F_{PR,\zeta} = -\frac{\beta (1-\mu)}{cr_1^2} \Bigg[\frac{\big[(\xi-\xi_1)\dot{\xi}+(\eta-\eta_1)\dot{\eta}+(\zeta-\zeta_1)\dot{\zeta}\big](\zeta-\zeta_1)}{r_1^2} + \dot{\zeta}\Bigg]\\
        \\
    \end{split}
    \label{4}
\end{equation}
We write the equations in a rotating frame 
$(x,y,z)$ by using the following transformation:
\begin{equation}
\begin{split}
  \xi &= xcos(t) + ysin(t)\\
  \eta &= xsin(t) - ycos(t)\\
  \zeta &= z  
\end{split}
\label{5}
\end{equation}
Here $(x,y,z)$ are chosen such that mean motion '$n$' is 1. 
Using the above transformation in eq.~(\ref{3}) and on simplification we get:
 
\begin{equation}
   \begin{split}
      \Ddot{x}-2\dot{y} & = x -\frac{(1-\beta)(1-\mu)(x+\mu)}{r_1^3} -\frac{a\mu(x-1+\mu)}{r_2^3} + (1+sw) F_{PR,x}
      \\
      \Ddot{y}+2\dot{x} & = y -\frac{(1-\beta)(1-\mu)y}{r_1^3} -\frac{a\mu y}{r_2^3} + (1+sw) F_{PR,y}
       \\
        \Ddot{z} & =  -\frac{(1-\beta)(1-\mu)z}{r_1^3} -\frac{a\mu z}{r_2^3} + (1+sw) F_{PR,z}\\
    \end{split}
    \label{6}
\end{equation} 
where,
\begin{equation}
    \begin{split}
 F_{PR,x} & = -\frac{\beta (1-\mu)}{cr_1^2} \Bigg[\frac{[(x+\mu)\dot{x}+(\dot{y}- \mu)y+z\dot{z}](x+\mu)}{r_1^2}+(\dot{x}-y)\Bigg]
 \\F_{PR,y} & = -\frac{\beta (1-\mu)}{cr_1^2} \Bigg[\frac{[(x+\mu)\dot{x}+(\dot{y}- \mu)y+z\dot{z}](y)}{r_1^2}+(x+\dot{y})\Bigg]
 \\F_{PR,z} & = -\frac{\beta (1-\mu)}{cr_1^2} \Bigg[\frac{[(x+\mu)\dot{x}+(\dot{y}- \mu)y+z\dot{z}](z)}{r_1^2}+(\dot{z})\Bigg]\\
   \end{split}
    \label{7}
 \end{equation}
 
 Now if we multiply first equation of system of equations(\ref{6}) by $\dot{x}$ second by $\dot{y}$ and third by $\dot{z}$ and adding and integrating them accordingly we get,

\begin{equation}
        \dot{x}^2+\dot{y}^2+\dot{z}^2= 2U-C 
        \label{8}
\end{equation}

where\\

\begin{equation}
U= \frac{(1-\beta)(1-\mu)}{r_1}+\frac{a\mu}{r_2}+\frac{1}{2} (x^2+y^2),
\label{9}
\end{equation}

and

\begin{equation}
C= \frac{2(1+sw)(1-\mu)\beta}{cr_1^2}\Bigg[\frac{(x\dot{x}+y\dot{y}+z\dot{z})^2}{r_1^2}- [(\dot{x}^2+\dot{y}^2+\dot{z}^2)-(\dot{x}y-x\dot{y})]\Bigg]
\label{10}
\end{equation}

From eq(\ref{7}),\\

$$\dot{x}^2+\dot{y}^2+\dot{z}^2=v^2$$\\
$v^2$ is the square of the velocity of the particle in the rotating frame.
Hence $$ v^2=2U-C$$
where $U$ is the potential  and $C$ is the Jacobi's constant.  The Jacobi's constant given in eq(\ref{10}), which we derive is the only constant of motion. 

For $v=0$, \\
$$2U-C=0$$ describes the Equipotential or Zero Velocity Curves.

\section{Collinear Lagrange Equilibrium Points and Zero Velocity Curves}

Lagrange equilibrium points are positions in space where all forces cancel each other. Location of Lagrange equilibrium points can be found by solving eq.~(\ref{6}) presuming that all derivatives of x,y,z  with respect to time are zero.\\
With this eq.~(\ref{6}) becomes,\\

\begin{equation}
        x-\frac{(1-\beta)(1-\mu)(x+\mu)}{r_1^3}-\frac{a\mu(x-1+\mu)}{r_2^3}-\frac{(1+sw)\beta(1-\mu)y}{cr_1^2}\Bigg[\frac{-\mu (x+\mu)}{r_1^2}-1\Bigg]=0\\
 \label{11}
\end{equation}

\begin{equation}
        y-\frac{(1-\beta)(1-\mu)y}{r_1^3}-\frac{a\mu y}{r_2^3}-\frac{(1+sw)\beta(1-\mu)}{cr_1^2}\Bigg[\frac{-\mu y^2}{r_1^2}+x\Bigg]=0\\
 \label{12}
\end{equation}

\begin{equation}
        z\Bigg[\frac{(1-\beta)(1-\mu)}{r_1^3}+\frac{a\mu }{r_2^3}+\frac{(1+sw)\beta(1-\mu)\mu y}{cr_1^4}\Bigg]=0\\
    \label{13}
\end{equation}
\noindent
From eq.(\ref{13}) we get two cases: \\

case (i)  $z=0$ \\

case (ii) $ \frac{(1-\beta)(1-\mu)}{r_1^3}+\frac{a\mu}{r_2^3}+\frac{(1+sw)\beta(1-\mu)\mu y}{cr_1^4}=0$\\

 In the classical case, there are only planar solutions $z= 0$ and we obtain the planar Lagrangian points,  which lie in orbital plane of the primaries \cite{moulton1914introduction}. In this case, we have out of plane solutions, 
 $z\neq 0$, i.e., out-of-plane equilibrium points.
\\

If we take $y=0$ and $z=0$ we obtain three collinear Lagrange points, that lie on the axis joining the two primary bodies hence. In this case eq.~(\ref{11}) becomes:

\begin{equation}
x-\frac{(1-\beta)(1-\mu)}{(x+\mu)^2}-\frac{a\mu}{(x-1+\mu)^2}=0
\end{equation}

This gives us a 5th degree polynomial, the real roots of which are the CLEP, which are denoted by  $L_1, L_2, L_3$. $\beta$ is the radiation pressure parameter. We have studied  the variation in the location of CLEP w.r.t change in  $\beta$, in the 0.1 to 0.9.  We have fixed the values of  $\mu$=0.1 and $a$= 0.8 since these are the most typical values  for planetary systems. We have assumed $sw$=0.35 as it is  the average value taken for the ratio of solar wind drag and PR drag \citep{gust94}. 
}. For $sw=0.35$, $\beta=0.1$, oblateness factor $a=0.8$, mass parameter $\mu=0.01$ the positions of the CLEP are given in the table \ref{T1}. 
\begin{table}[ht]
    \centering
    \begin{tabular}{||c| c |c| c||}
    \hline
    LP  &  X & Y & C  \\[0.5ex]
    \hline
    L1   & -0.8443   & 0 &   2.8575  \\
    L2   & -1.1256   & 0 &   2.8719  \\
    L3   &  0.9696   & 0 &   3.5446 \\ [1ex]   
    \hline
    \end{tabular}
    \caption{Location of the Collinear Lagrange Equilibrium points.}
    \label{T1}
\end{table}

Zero velocity curves are plotted for different values of mass ratio, radiation pressure parameter, and oblateness factor

\begin{figure}[H]
\scalebox{0.6}[0.6]{\includegraphics[width=\columnwidth]{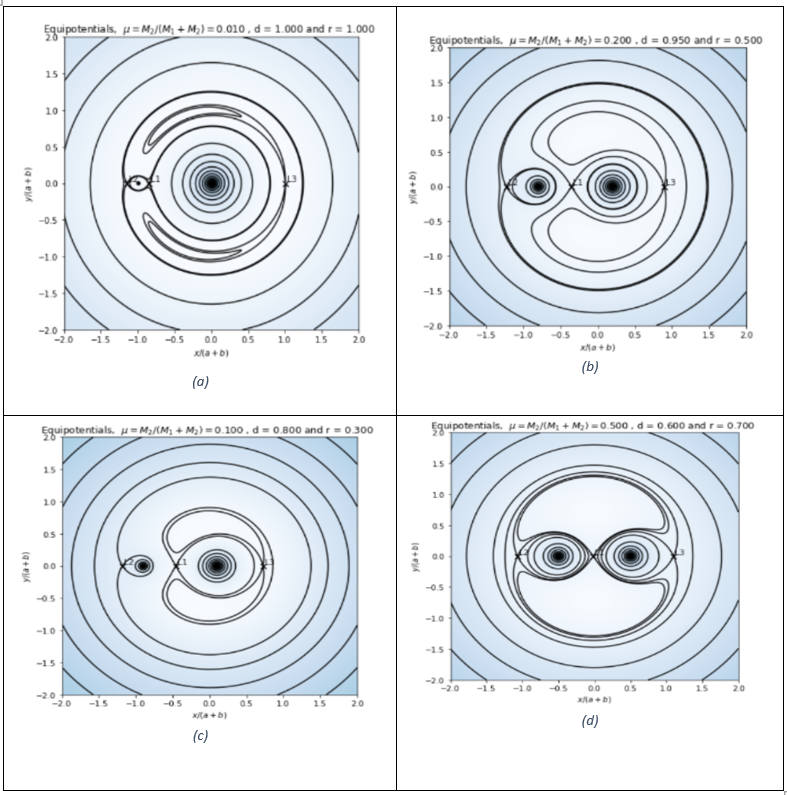}}\centering
    \caption{Location of Lagrangian points are shown in the figures for different parameters values \\(a)  $\mu=0.01$, $\beta=1$, $a=1$, \\ (b) $\mu=0.2$, $\beta=0.5$, $a=0.95$, \\ (c) $\mu=0.1$, $\beta=0.3$, $a=0.8$ , \\ (d)  $\mu=0.5$, $\beta=0.7$, $a=0.6$}
    \label{F2}
\end{figure}

\section{Variations of Lagrange equilibrium points}
As we see that locations of the Lagrange equilibrium points depends on the values of radiation pressure parameters $\beta$, mass ratio $\mu$ and oblateness factor $a$ hence, these positions changes with changes in the parameter values.

\subsection{Variations in CLEPs with Radiation pressure.}
Stars and planets in their early stage of formation shows drastic changes in radiations, hence changes the positions of the equilibrium points.\\
To study the shift in the locations we change the radiation pressure parameter $\beta$ from $0.1$ to $0.9$ and the values of mass ratio $\mu$ and oblateness factor $a$ are kept fixed at $0.1$ and $0.8$ respectively. \\
As radiation of the body increases the CLEPs move away the more massive radiating body.

\subsection{Variations in CLEP with Oblateness.}
Rotation of the planet induces them to distort their shapes from the perfect sphere. To study the variation we reduce the oblateness of the body by varying $0.1 \leq a \leq 1$, and take the value of radiation pressure parameter $\beta=0.6$ and mass ratio $\mu = 0.1$.
We see that as oblateness increases i.e. value of the $a$ decreases, the location of $L_1, L_2, L_3$ moves away from the oblate body.

\begin{figure}[H]
\scalebox{0.4}[0.4]{\includegraphics{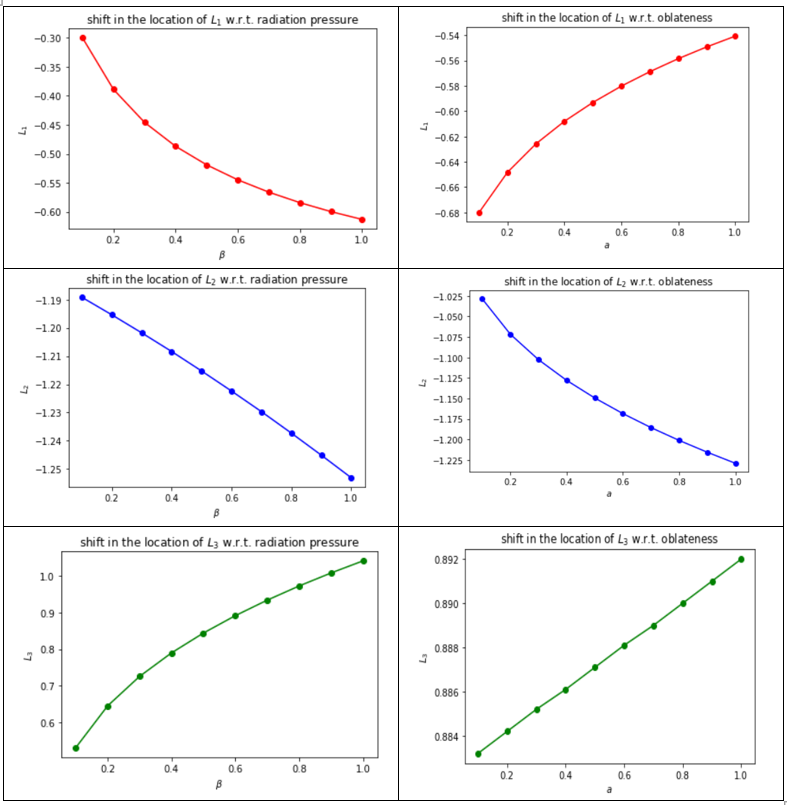}}\centering
    \caption{Variations in the Locations of CLEPs  \\ Fig.(a) shows the variation in CLEPs with changes in the radiation pressure $\beta$, red dot shows location for $\beta= 0.1$, blue dot is for  $\beta=0.2$, green is for $\beta= 0.3$ and so on... And keeping $\mu = 0.1$ and oblateness factor $a=0.8$ fixed.\\  Fig.(b)  shows shows the variation in CLEPs with changes in the oblateness $a$, red dot shows location for $a=0.1$, blue dot is for  $a=0.2$, green is for $a=0.3$ and so on...  and keeping $\mu = 0.1$, $\beta=0.6$ fixed.}
    \label{F3}
\end{figure}

\begin{figure}[H]
\scalebox{0.6}[0.6]{\includegraphics{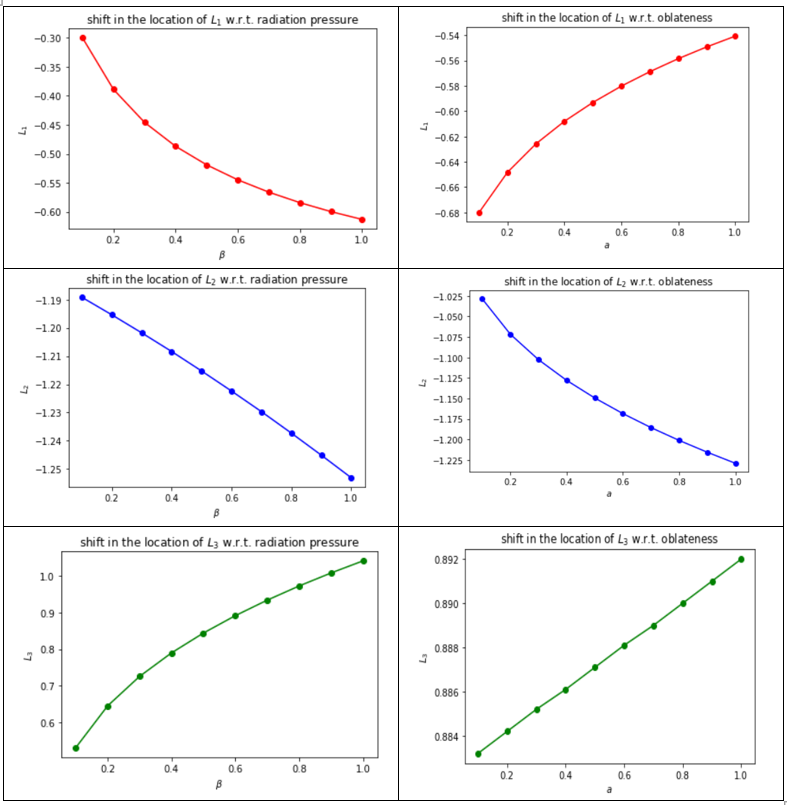}}\centering
    \caption{The figures on the right hand side shows the shift in the Lagrange points with varying oblateness parameter from $0.1$ to $1$ and the figures on the left hand side shows the shift in the Lagrange points with varying the radiation pressure parameter from $0.1$ to $1$  }
\label{F4}   
\end{figure}

\section{Stability of Lagrange points}
In this section we study the stability of the CLEP  which is Lyapunov stability. This is because we are interested in investigating the possibility of matter accretion near the Lagrangian points and  remain the vicinity for a long enough period to form planets or minor objects. 

We take small displacements $\xi$ and $\eta$ in the $x$ and $y$ directions respectively and substitute them in the equation of motion and on simplifying we obtain:

$$\ddot{\xi}-2\dot{\eta}=\xi U_{xx} +\eta U_{xy}$$
$$\ddot{\eta}+2\dot{\xi}=\xi U_{yx} +\eta U_{yy}$$
We get the characteristic equation of the above system as

\begin{center}
\begin{equation}
\lambda^4 + \alpha_1 \lambda^2 +\alpha_2=0
\label{lambda}
\end{equation}
\end{center}

where,\\
$\alpha_1= 4-U_{xx}-U_{yy}$,  \\
$\alpha_2 = U_{xx} U_{yy}-(U_{xy})^2$\\
and 
\begin{equation}
    \begin{aligned}  
         U_{xx} & =  1-(1-\beta)(1-\mu)\Bigg[\frac{1}{[(x_0+\mu)^2+y_0^2]^{3/2}}-             
        \frac{3(x_0+\mu)^2}{[(x_0+\mu)^2+y_0^2]^{5/2}}\Bigg]\\
        &-(a\mu)\Bigg[\frac{1}{[(x_0+\mu-1)^2+y_0^2]^{3/2}}-\frac{3(x_0+\mu-1)^2}{[(x_0+\mu-1)^2+y_0^2]^{5/2}}  \\      
    \end{aligned} 
    \label{14}
\end{equation}

\begin{equation}
    \begin{aligned}
        U_{yy} & =   1-(1-\beta)(1-\mu)\Bigg[\frac{1}{[(x_0+\mu)^2+y_0^2]^{3/2}}-\frac{3y_0^2}{[(x_0+\mu)^2+y_0^2]^{5/2}}\Bigg]\\
        &-(a\mu)\Bigg[\frac{1}{[(x_0+\mu-1)^2+y_0^2]^{3/2}}-\frac{3y_0^2}{[(x_0+\mu-1)^2+y_0^2]^{5/2}} \\ 
    \end{aligned}
    \label{15}
\end{equation}

\begin{equation}
    \begin{aligned}
        U_{xy} & =  \frac{3(1-\beta)(1-\mu)y_0(x_0+\mu)}{[(x_0+\mu)^2+y_0^2]^{5/2}}+\frac{3a\mu y_0(x_0+\mu -1)}{[(x_0+\mu -1)^2+y_0^2]^{5/2}}\\  
    \end{aligned}
    \label{16}
\end{equation} \\

For the Collinear Lagrange Points, we take $y=z=0$ in Equations \ref{14} -\ref{16} which gives
\begin{equation}
 U_{xx} = 1- 2\Bigg[\frac{(1-\beta)(1-\mu)}{(x+\mu)^3} + \frac{a\mu}{(x+\mu-1)^3}\Bigg] = 1 - 2f \\
 \label{18}
\end{equation}
\begin{equation}
 U_{yy} = 1- \Bigg[\frac{(1-\beta)(1-\mu)}{(x+\mu)^3} + \frac{a\mu}{(x+\mu-1)^3}\Bigg] = 1-f \\
 \label{19}
\end{equation}
\begin{equation}
 U_{xy} = 0
 \label{20}
\end{equation}
Where,   
\begin{equation}
\begin{aligned}
f = \Bigg[\frac{(1-\beta)(1-\mu)}{(x+\mu)^3} + \frac{a\mu}{(x+\mu-1)^3}\Bigg]
\label{stable}
\end{aligned}
\end{equation}
   
hence $$ \alpha_1= 2+3f$$     and,
$$\alpha_2= 1-3f+2f^2$$
After putting all these in the characteristic eq.(\ref{lambda}) the condition for stable CLEP's (i.e. Imaginary values for all the eigen values) will be $$f>1$$
i.e. 
\begin{equation}
\Bigg[\frac{(1-\beta)(1-\mu)}{(x+\mu)^3} + \frac{a\mu}{(x+\mu-1)^3}\Bigg] > 1
\label{cond}
\end{equation}

The stability of the Lagrangian points is obtained from the roots of the characteristic equation \ref{lambda}. For Lagrangian points to be stable all the eigenvalues of the characteristic equation need to be purely imaginary.  If and only if all the roots are purely imaginary then the Lagrangian points will be stable in Lyapunov sense, since only purely imaginary roots will give a periodic solution and the infinitesimal mass will remain in the vicinity of the Lagrangian point for a sufficiently long time. On the other hand complex or purely real roots will give unstable Lagrange points. The condition of stability for the collinear points is given in Equation \ref{cond}. \\

Since the $x$ coordinate of $L_1$ and $L_2$ is always negative, and that of $L_3$ is positive, we note from Equation~\ref{cond} that the condition for stability can be satisfied only for $L_3$. Hence the Lagrangian point $L_3$ is conditionally stable.

\section{Application to Exoplanetary Systems}
In this section, we discuss the stability of CLEPs for the sample of ten exoplanet systems for which  data has been taken from the NASA Exoplanet Data Archive as shown in Table \ref{T2}.
To obtain the Lagrangian points, we selected systems which has atleast one planet around a star. We also needed to find the radiation parameter, i.e., the luminosity of the star as well as oblateness and masses. The sample was selected for exoplanets with varying value of radiation pressure from host stars with  temperatures between 3500K to 6500K ( K, G and F-type stars).

\begin{table}[H]
    \centering
    
    \begin{tabular}{||l|c|c|c||}
    \hline
    Exoplanet systems  &  $\beta$ & $\mu$ & $a$ \\[0.5ex]
    \hline
    CoRoT-2b        & 0.0155   & 0.0032  & 0.9997   \\
    TOI-1278 b      & 0.0023   & 0.0318  & 0.9999   \\
    HAT-P-20 b      & 0.0052   & 0.0093  & 0.9999   \\
    Kepler-75 b     & 0.0118   & 0.0105  & 0.8214    \\
    WASP- 89 b      & 0.0097   & 0.0060  & 0.9940    \\
    TIC 172900988 b & 0.0381   & 0.0020  & 0.9966    \\
    NGTS 9b         & 0.0422   & 0.0021  & 0.9982   \\
    LP 714 - 47 b   & 0.0026   & 0.0002  & 0.9992     \\
    WASP- 162 b     & 0.0189   & 0.0052  & 0.9009    \\
    XO- 3 b         & 0.0498   & 0.0092  & 0.9656     \\
    \hline
    \end{tabular}
    \caption{Numerical Data}
    \label{T2}
\end{table}

We selected a sample of exoplanet systems with a single planet from the NASA Exoplanet Data Archive shown in Table \ref{T2}. We calculated the oblateness factor $a$ from eccentricity by,

$$a= \sqrt{1-(e^2)}$$

and the radiation pressure parameter $\beta$, from Stephan-Boltzman's law is,\\
$$\beta = \frac{3 \sigma R^2T^4}{2GMr \rho }$$  \\
Where $R$, $T$, $M$ are stellar radius, stellar effective temperature and stellar mass respectively, and $G$ is the gravitational force constant $G= 6.67 \times 10^-11 \rm{m^3/kg^2}$, $\sigma = 5.67 \times 10^ -8 \rm{W/m^2K^4}$ is the Boltzman's constant. The radius and density of the dust particle in the system is $r = 2 \times 10^{-2}\rm{cm}$ and $\rho = 1.4 \rm{gm/cm^3}$  respectively. \\

Characterstic roots for the data from \ref{T2} have been calculated for CLEPs $L_1$,$L_2$ and $L_3$  and are listed in  Table \ref{T3}--\ref{T5}.

\begin{table}[H]
    \centering
    \caption{Characterstic roots for $L_1$ point}
    \begin{tabular}{||c| c |c| c||}
    \hline
    Exoplanet systems  &  $L_1$ & $\pm \lambda_{1,2}$ & $\pm \lambda_{3,4}$ \\[0.5ex]
    \hline
    CoRoT-2b        & $(-0.8957,0)$ & $0.9824$ & $2.0847\it{i}$ \\
    TOI-1278 b      & $(-0.7639,0)$ & $1.4432$ & $2.5594\it{i}$ \\
    HAT-P-20 b      & $(-0.8516,0)$ & $1.1161$ & $2.21248it{i}$ \\
    Kepler-75 b     & $(-0.8525,0)$ & $1.1111$ & $2.2078\it{i}$ \\
    WASP- 89 b      & $(-0.8724,0)$ & $1.0498$ & $2.1480\it{i}$ \\
    TIC 172900988 b & $(-0.9086,0)$ & $0.9319$ & $2.0389\it{i}$  \\
    NGTS 9b         & $(-0.9069,0)$ & $0.9327$ & $2.0397\it{i}$\\
    LP 714 - 47 b   & $(-0.9626,0)$ & $0.8298$ & $1.9512\it{i}$ \\
    WASP- 162 b     & $(-0.8808,0)$ & $1.0207$ & $2.1204\it{i}$\\
    XO- 3 b         & $(-0.8475,0)$ & $1.0923$ & $2.1893\it{i}$ \\
    \hline
    \end{tabular}
    \label{T3}
\end{table}

\begin{table}[H]
    \centering
    \caption{Characterstic roots for $L_2$ point}
    \begin{tabular}{||c| c |c| c||}
    \hline
    Exoplanet systems  &  $L_2$ & $\pm \lambda_{1,2}$ & $\pm \lambda_{3,4}$ \\[0.5ex]
    \hline
    CoRoT-2b        & (-1.1013,0) & 0.5211 & 1.737832\it{i}  \\
    TOI-1278 b      & (-1.2040,0) & 0.3478 & 1.650364\it{i} \\
    HAT-P-20 b      & (-1.1428,0) & 0.4545 & 1.698284\it{i} \\
    Kepler-75 b     & (-1.1373,0) & 0.4628 & 1.702496\it{i} \\
    WASP- 89 b      & (-1.1240,0) & 0.4887 & 1.716077\it{i} \\
    TIC 172900988 b & (-1.0839,0) & 0.5451 & 1.747966\it{i} \\
    NGTS 9b         & (-1.0846,0) & 0.5405 & 1.745227\it{i} \\
    LP 714 - 47 b   & (-1.0374,0) & 0.6680 & 1.827509\it{i} \\
    WASP- 162 b     & (-1.1135,0) & 0.5016 & 1.723090\it{i} \\
    XO- 3 b         & (-1.1366,0) & 0.4292 & 1.685871\it{i} \\
    \hline
    \end{tabular}
    \label{T4}
\end{table}

\begin{table}[H]
    \centering
    \caption{Characterstic roots for $L_3$ point}
    \begin{tabular}{||c| c |c| c||}
    \hline
    Exoplanet systems  &  $L_3$ & $\lambda_{1,2}$ & $ \lambda_{3,4}$ \\
    %[0.5ex]
    \hline
    CoRoT-2b     & (0.9962,0) & $\pm 4116$ & $\pm 5836$\\
    TOI-1278 b      & (1.0125,0) & $\pm 19.037$ \it{i}  & $\pm 27.231$ \it{i}   \\
    HAT-P-20 b      & (1.0021,0)  & $\pm78.975$\it{i} & $\pm111.76$\it{i}   \\
    Kepler-75 b     & (1.0003,0)  & $\pm82.518$\it{i}  & $\pm116.771$\it{i}  \\
    WASP- 89 b    & 0.9992,0)   & $\pm 196.43$\it{i}  & $\pm 289.32$\it{i} \\
    TIC 172900988 b & (0.9880,0)   & $\pm 44.539$  & $\pm 62.852$\\
    NGTS 9b         & (0.9866,0)   & $\pm 37.681$  & $\pm 53.128$   \\
    LP 714 - 47 b   & (0.9992,0)   & $1110.13$ & $784.261 $   \\
    WASP- 162 b     & (0.9958,0)   & $\pm 0.00111$ \it{i}  & $\pm 3721.4 $ \it{i}    \\
    XO- 3 b         & (0.9870,0)   & $\pm 414.1784$  & $\pm 585.51$ \\
    \hline
    \end{tabular}
    \label{T5}       
\end{table}

From Tables \ref{T3}--\ref{T5} we notice that for none of the exoplanetary systems, all eigenvalues for $L_1$ and $L_2$ are purely imaginary, hence we conclude that $L_1$ and $L_2$ for all the systems under investigation are unstable. For some exoplanetary systems in consideration, the eigenvalues of the characteristic equation for $L_3$ have purely imaginary roots for $0.002< \beta < 0.02$ and $0.005 < \mu < 0.03$ which makes $L_3$ conditionally stable.

\section{Conclusion}
We study the RTBP under the influence of perturbations like radiation pressure, P--R drag, and oblateness and find the location of the  three CLEPs and we observe that the locations vary by altering the parameters {\it{viz.}}-- $\mu$ mass of the primary,  $\beta$-- the radiation parameter,  and  $a$ --the oblateness parameter. Further, we study the stability of these Lagrange points. We apply our results to examine the location and stability of ten exoplanet systems namely  CoRoT-2b, TOI-1278 b,  HAT-P-20 b, Kepler-75 b,  WASP- 89 b, TIC 172900988 b, 
 NGTS 9b,  LP 714 - 47 b,  WASP- 162 b and  XO- 3 b. For the ten exoplanet systems we have investigated,  we note that $L_1$ and $L_2$ are always unstable while $L_3$ becomes stable if the value of $\beta$ is in between $0.002$ to $0.02$  and $\mu$ is between $0.005$ to $0.03$. 

Column 2 of Tables 3--5 shows the locations of the Lagrange points for the exoplanet systems. These are the locations one can expect to find minor bodies, debris or planetesimals - which may act as seeds for planet formation. We can see from Figure \ref{F4} that the effect of variation of the radiation pressure is an order of magnitude greater than the effect of variation of the oblateness on the location of the Lagrange points. We also notice that the location of the Lagrange point shifts away from the radiating body as the radiation increases and the Lagrange points shift away from the oblate body as the oblateness decreases.

From the theory of star formation and evolution, we know that in the initial stages, a proto-star as it moves along the Henyey track and arrives on the Main sequence, it goes through pulsation, and consequently its radiation varies. With a change in radiation pressure we notice that there is a change in the location of the Lagrange Points hence this could be one of the mechanisms for migration of planets to other locations from the point of formation.

This work is important in locating points in protoplanetary systems where planet formation can commence. 

\section*{Acknowledgements} This research has made use of the NASA Exoplanet Archive, which is operated by the California Institute of Technology, under contract with the National Aeronautics and Space Administration under the Exoplanet Exploration Program.

%%use \balance somewhere in the left column of the last page to balance the two columns in the end page

%%References section
\bibliography{Myref.bib}
\vspace{-1.5em}
\end{document}